\documentclass[sigconf, screen=true]{acmart}

\usepackage{todonotes}
\usepackage{hyperref}

\usepackage{acronym}
\AtBeginDocument{%
  \providecommand\BibTeX{{%
    \normalfont B\kern-0.5em{\scshape i\kern-0.25em b}\kern-0.8em\TeX}}}

\copyrightyear{2020}
\acmYear{2020}
\setcopyright{acmlicensed}\acmConference[IVA '20]{IVA '20: Proceedings of the 20th ACM International Conference on Intelligent Virtual Agents}{October 19--23, 2020}{Virtual Event, Scotland Uk}
\acmBooktitle{IVA '20: Proceedings of the 20th ACM International Conference on Intelligent Virtual Agents (IVA '20), October 19--23, 2020, Virtual Event, Scotland Uk}
\acmPrice{15.00}
\acmDOI{10.1145/3383652.3423860}
\acmISBN{978-1-4503-7586-3/20/09}

\newacro{AMT}{Amazon Mechanical Turk}
\newacro{AVAC}[AV attention check]{audio-video based attention check}
\newacro{SVAC}[SV attention check]{same video attention check}
\newacro{ICC}{intraclass correlation coefficients}
\newacro{IMC}{instructional manipulation check}

\usepackage[]{algorithm2e}

\begin{document}

% \listoftodos{}

%\title{Can we trust online crowd-workers?\\Comparing online and offline participants for an audio-visual perception study}
\title{Can we trust online crowdworkers?\\Comparing online and offline participants in a preference test of virtual agents}

\author{Patrik Jonell*}
\affiliation{%
  \institution{KTH Royal Institute of Technology}
 }
 \email{pjjonell@kth.se}

 \author{Taras Kucherenko*}
 \affiliation{\institution{KTH Royal Institute of Technology}}
 \email{tarask@kth.se}

 \author{Ilaria Torre}
 \affiliation{\institution{KTH Royal Institute of Technology}}
 \email{ilariat@kth.se}

 \author{Jonas Beskow}
 \affiliation{\institution{KTH Royal Institute of Technology}}
 \email{beskow@kth.se}
 
\thanks{*Both authors contributed equally to this research}

% Inviting participants to the lab has always been a time-consuming and, frankly speaking, an inefficient endeavor. Some user studies, such as surveys and perception studies, can be run both in-lab and online. 
% we might want to add that sometimes it is actually impossible 

\begin{abstract}
Conducting user studies is a crucial component in many scientific fields. While some studies require  participants to be physically present, other studies can be conducted both physically (e.g. in-lab) and online (e.g. via crowdsourcing). Inviting participants to the lab can be a time-consuming and logistically difficult endeavor, not to mention that sometimes research groups might not be able to run in-lab experiments, because of, for example, a pandemic. Crowdsourcing platforms such as \ac{AMT} or Prolific can therefore be a suitable alternative to run certain experiments, such as evaluating virtual agents. Although previous studies investigated the use of crowdsourcing platforms for running experiments, there is still uncertainty as to whether the results are reliable for perceptual studies. Here we replicate a previous experiment where participants evaluated a gesture generation model for virtual agents. The experiment is conducted across three participant pools -- in-lab, Prolific, and \ac{AMT} -- having similar demographics across the in-lab participants and the Prolific platform. Our results show no difference between the three participant pools in regards to their evaluations of the gesture generation models and their reliability scores. The results indicate that online platforms can successfully be used for perceptual evaluations of this kind.
\end{abstract}

\begin{CCSXML}
<ccs2012>
   <concept>
       <concept_id>10003120.10003121.10011748</concept_id>
       <concept_desc>Human-centered computing~Empirical studies in HCI</concept_desc>
       <concept_significance>500</concept_significance>
       </concept>
   <concept>
       <concept_id>10003120.10003121.10003122.10003334</concept_id>
       <concept_desc>Human-centered computing~User studies</concept_desc>
       <concept_significance>300</concept_significance>
       </concept>
 </ccs2012>
\end{CCSXML}

\ccsdesc[500]{Human-centered computing~Empirical studies in HCI}
\ccsdesc[300]{Human-centered computing~User studies}


% \keywords{evaluation, user study, online workers, virtual agents, Prolific, AMT}


\maketitle

\section{Introduction}

More and more perceptual studies in the Human-Computer Interaction field are done using online crowdsourcing platforms, such as \acl{AMT}\footnote{www.mturk.com} (\acs{AMT}) and Prolific\footnote{www.prolific.co} ~\cite{sadoughi2019speech, yoon2019robots}. As there is no way to control the environment and experimental setting of the online workers, they can do other activities simultaneously or ignore instructions (such as wearing headphones). This can in turn lead to poor quality of study results. Checking attentiveness of online workers is common practice for perceptual studies~\cite{fleischer2015inattentive, cheung2017amazon} and often leads to discarding a large number of participants~\cite{jonell2019learning, kucherenko2019analyzing}. Passing attention checks does not always imply good concentration however, since online workers can simply learn to just pass the attention checks~\cite{lovett2018data}. 
%Hence it remains unknown whether attention checks is a reliable way of assessing online workers' performance when participating in crowsourced user studies.

One way to investigate reliability of online participants is to compare them with in-lab (offline) participants. Offline participants are believed to be more attentive because they are in a controlled environment with fewer distractions and often with an experimenter present in the same room~\cite{naderi2015effect}.

%User studies can be divided in two categories: those which are purely based on text and those which involve another stimuli as well.
Prior work has been investigating differences between online and offline study participants \cite{hauser2016attentive, kees2017analysis, crump2013evaluating, byun2015online, mcnaney2016speeching}, but most of them compared text-based survey research (questionnaire studies). In this work we focus on perceptual studies which differ from questionnaire studies in that they involve stimuli of other modalities than text. While it might be straightforward to prove attentiveness and reliability for text-based studies, the same is not always the case for perceptual studies, where there might not be a right or wrong answer to a question.
%Perceptual studies are arguably more subjective and thus makes it harder to evaluate the reliability of the provided answers. 
%More in-depth discussion on screening is provided in Section \ref{subs:screening}.

In this paper we replicate the study conducted by Kucherenko et al. ~\cite{kucherenko2020gesticulator} to investigate the differences in performing subjective perceptual studies between an in-lab setting and two crowdsourcing platforms. Specifically, we consider a preference test between two gesture generating models for a virtual agent where video artifacts have been produced for both models. The study was repeated three times: in-lab, using \ac{AMT}, and on Prolific. In order to compare %evaluate the differences between 
the participants in the three different pools, we evaluate the difference in the preference score given to the two gesture generation models, the difference in inter- and intra-rater agreement, and the number of attention checks passed.

Our main research question is: \textit{Do in-lab participants perform differently from participants on crowdsourcing platforms in a subjective audio-visual preference experiment?}
% Is there a difference between running an experiment investigating subjective perceptual differences in video-pairs using a crowd-sourcing platforms vs. an in-lab study? 

% \subsection*{Pre-registration}
 % https://osf.io/dxwak

% The contributions of our paper are the following:
% \begin{enumerate}
%     \item Comparison of in-lab and offline participants with the similar demographics for the audio-visual perception task
%     \item Reproducing previous study \cite{kucherenko2020gesticulator} for new study participants pools: not only commonly used AMT, but also in-lab and online at Prolific
%     %\item Comparison of different tools: AMT and Prolific Academic
%     %\item Analysis of attention checks as filtering for the quality of the data in audio-visual perception task
    
% \end{enumerate}

\section{Background and Related Work}

In this section, we review previous work on analysing and comparing the quality of the data obtained from online workers with that from in-lab participants.

%Since researchers in various fields are increasingly turning to online data collection panels for research purposes, there is a substantial body of research on analysing and comparing the quality of the data obtained from online workers with that from the in-lab participants.

%Improving the quality of the data obtained from online workers is also an active field of research ~\cite{maniaci2014caring, berinsky2016can, desimone2018dirty}. Several experiments indicated that data screening can have a moderate to large impact on statistical results \cite{maniaci2014caring, desimone2018dirty}. Berinsky et al.~\cite{berinsky2016can} examined several strategies to enforce workers to be more attentive on different tasks. They have found that some of them (especially training workers) produced a strong effect on the IMC passage rate, but that did not translate into higher-quality data. It has been shown that online workers are on the lookout for attention checks \cite{lovett2018data}. In other words, online workers just learn to pass tests. 

% germine2012web - no filtering
% byun2015online experts vs mturkers. They use filtering. Short one utterance audio clips to listen to
% lansford2016use - no filtering
% mcnaney2016speeching - not really for studies. It uses crowd workers for assesment/labeling of data
% Komarov:2013:CPE:2470654.2470684 - workers evaluate user interfaces between in-lab and mturk

\subsection{Improving quality of online studies}
Improving the quality of the data obtained from online workers is an active field of research ~\cite{maniaci2014caring, berinsky2016can, desimone2018dirty}. One way of detecting participants who might not be paying attention is to use \acp{IMC}. These were first introduced by Oppenheimer et al. \cite{oppenheimer2009instructional}, and are one of the most common ways to detect "cheaters" or inattentive participants. As the name suggests, \acp{IMC} are manipulations of the instructions which are used to detect if participants read the instructions carefully. An \ac{IMC} could, for example, be an instruction which tells the participant to ignore a specific question, click "other" or write "I read the instruction" as an answer. Since \acp{IMC} were first introduced there have been other methods developed to detect inattentive participants or "dirty data" (reviewed in \cite{curran2016methods}), but using \acp{IMC} is still the standard technique.

Berinsky et al.~\cite{berinsky2016can} examined several strategies to enforce workers to be more attentive on different tasks. They found that some of them (especially training workers) produced a strong effect on the \ac{IMC} passage rate, but that did not translate into higher-quality data. Apart from that, it has been shown that online workers are on the lookout for attention checks \cite{lovett2018data}.
%In other words, online workers just learn to pass the attention tests.

There are ongoing debates on whether attention checks should be used. Hauser et al.~\cite{hauser2018manipulation} argued that attention checks might distort the results, especially if they are very different from the original task. On the other hand, Kung et al.~\cite{kung2018attention} experimentally showed that common attention checks do not affect scale validity in several classical experiments. 
In other words, previous research suggests that attention checks could be used, but with care~\cite{cheung2017amazon}. %The attention checks that we developed for our study are similar to the actual task and should therefore not distort the results. 

%\subsection{Screening}
%\label{subs:screening}

Another way to ensure that participants provide high quality data is by screening them. This can be done by removing participants who are deemed to be unfit according to some criteria, such as by providing several wrong answers to questions with known answers~\cite{maniaci2014caring} or by giving many identical answers in a row~\cite{costa2008revised}. 
Several experiments indicated that screening can have an  %moderate to large
impact on statistical results \cite{maniaci2014caring, desimone2018dirty}. Screening therefore is commonly used for online studies, and can be done during or after the study.

%\subsection{Screening in gesture generation field}

\subsection{Comparing online and offline participants}

The type of data a study seeks to collect could also influence whether screening methodologies are more or less successful. For example, a qualitative study such as a market research, where participants have to create elaborate, free-text answers, might need different attention checks than an online questionnaire, where participants have to respond using multiple-choice answers.

Several researchers have been investigating differences between online and offline participants for questionnaire studies \cite{hauser2016attentive, kees2017analysis, crump2013evaluating}. Hauser and Schwarz~\cite{hauser2016attentive} used \acp{IMC} to test the attentiveness of participants when they read instructions before filling out questionnaires. For this study they used in-lab participants which were using their own computers and were not supervised. In three studies, \ac{AMT} workers were consistently more likely to pass \acp{IMC} than the in-lab participants. Kees et al~\cite{kees2017analysis} did a similar study but found no differences between \ac{AMT} and in-lab participants in terms of their performance in the tests.

% Questionnaires are, for instance, text-based and those studies focused on attentiveness during instruction reading. Thus, participants might have missed a lot of information by skipping some of the text they had to read.

% Other types of studies, such as those based on audio-video perception, might not be as constrained by text. In cases where participants need to watch a video or listen to some audio, their attentiveness can be checked by asking them to do something based on the content of that video, thus relying less on attentiveness while reading instructions.

%\subsection{Perceptual research}
Several studies have indicated that online participants can reproduce in-lab results for different perceptual studies \cite{germine2012web, byun2015online, lansford2016use, mcnaney2016speeching, Komarov:2013:CPE:2470654.2470684}. %, while having different demographics. They were considering tasks that are more simple and require less focus than watching a video since they contained just one modality such as evaluation of sound \cite{lansford2016use, byun2015online}, picture \cite{germine2012web}.  
Lansford et al.~\cite{lansford2016use} found similar results for the online and offline participants in terms of perceptual-training benefits while having different demographics.
Germine et al.~\cite{germine2012web} showed that for challenging cognitive and perceptual experiments, online participants perform similarly to in-lab participants in different cognitive ability tests, even when those self-selected online participants are anonymous, uncompensated, and unsupervised. For a detailed review of perceptual studies, we refer the reader to Woods et al.~\cite{woods2015conducting}. %None of these perceptual studies controlled for demographics, which is important, while we ensure that online workers and in-lab participants in our study have very similar demographic properties. 

%, such as Cambridge Memory Face Test or Reading the Mind in the Eyes,

One particularly interesting study is that of Burmenia et al.~\cite{burmania2015increasing}. They conducted a perceptual study on emotion annotation in videos using \ac{AMT}. They proposed and evaluated a novel filtering method which uses online quality assessment, stopping the evaluation when the quality of the worker drops below a threshold. %they also had tests looking similar to the actual task, so that it is less clear that it is a test
They did, however, not compare in-lab with online participants.

The most similar study to ours is that of Byun et al.~\cite{byun2015online}. They compared in-lab experts with crowdworkers on \ac{AMT} in a speech perception task. They had certain stimuli which were expected to yield a certain result and filtered out workers that did not score above chance at those tasks. The main difference to this work is the fact that we are using not only audio, but audio-visual stimuli with longer duration (10s). The current study also differs from Byon et al.'s in that we do not compare expert judgments, but rather layman judgments, and control for some of the demographic attributes when possible. Additionally we have a subjective task with no pre-defined ``correct'' answers.% is determined by the participants. 
%A benefit of a subjective task is that it is harder to share "correct" answers online, which might otherwise impair the results.
%Having a subjective task also makes it harder to share correct answers online, which can otherwise happen and impair the study results.
\begin{figure*}[ht]
  \centering
    \includegraphics[width=0.8\textwidth]{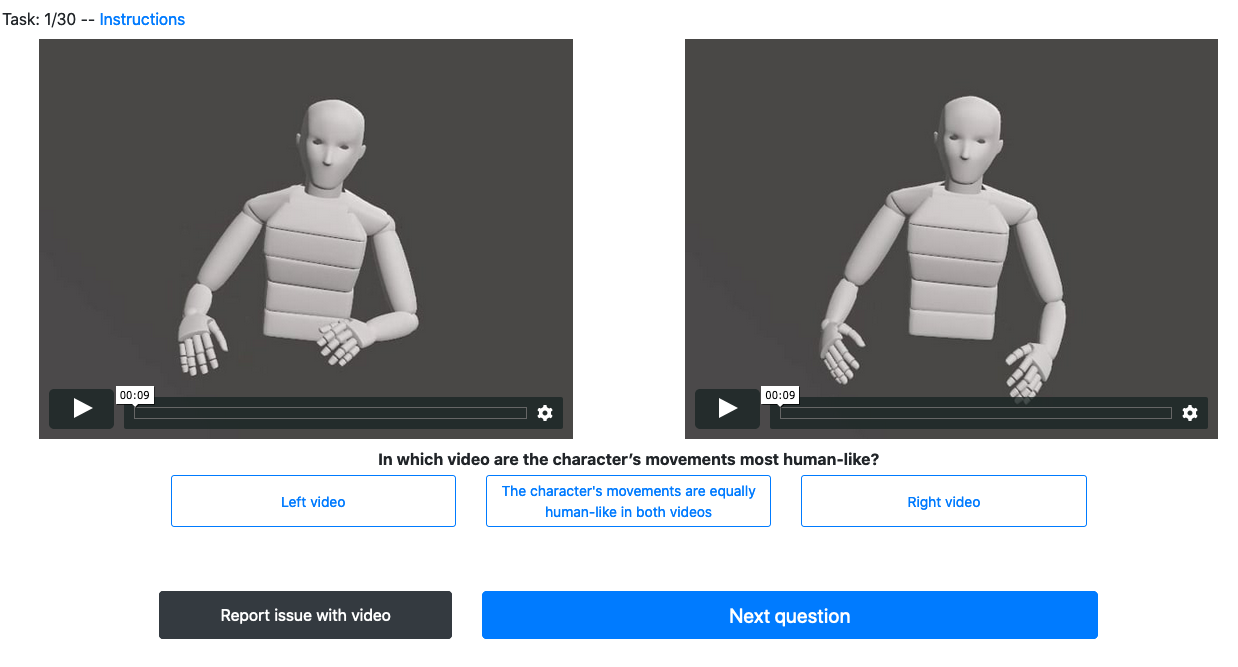}
  \caption{The user interface for evaluating the videos. Up in the left the participant is able to see how many trials are left. The two videos are played independently from each other. The participant has to chose one of the three alternatives as an answer in order to click ``Next question'' or click on the ``report video as broken'' button.}
  \label{fig:setting}
\end{figure*}

In the field of non-verbal behavior generation for virtual agents, subjective evaluation is required to assess the quality of the models. Most of the modern methods in this field conduct subjective evaluations using online crowdsourcing platforms, such as \ac{AMT} \cite{yoon2019robots, jonell2019learning, kucherenko2020gesticulator}. Many of them do screening based on attention checks. Yoon et al. \cite{yoon2019robots} excluded participants who could not pass attention check questions or gave too vague answers to questions about subjective impressions, resulting in the exclusion of 28\% of participants. Jonell et at. \cite{jonell2019learning} discarded 43\% of participants who did not pass the attention checks. Kucherenko et al. \cite{kucherenko2020gesticulator} used attention checks which were realized by distorting either the audio or the video. Participants who failed to report the majority of those samples as having an issue were discarded. Most of the participants (79\%) did not finish the experiments as they either dropped out or failed a majority of the attention checks. Those results put in question the reliability of online workers for audio-visual perception studies.

The present study investigates if online participants are as reliable as in-lab participants. To the best of our knowledge, this is the first study which makes this investigation using virtual agents.
%whilst evaluating its gestures.

%Sescleifer et al.~\cite{sescleifer2018systematic} overview the online perceptual studies to evaluate speech quality and conclude that they provide high-quality data as in-lab participants. 

% Byun et al.~\cite{byun2015online} compared AMT workers with experts for ratings of speech samples and have got the same results as experts with a few AMT workers. 

\section{Method}
The study used a mixed design with two independent variables: \linebreak a) participant pool (between-subjects: in-lab, Prolific, \ac{AMT}) and b) gesture generation model (within-subject: `No PCA' and `No text').

The main independent variable was the participant pool.
The participant pools vary in how much control they provide, with the in-lab study granting a higher degree of control, but over a limited set of participants with limited demographics. On the other side of the spectrum we find \ac{AMT}, which affords little control but a large amount of participants with a wide spread in demographics. There are also other services, such as Prolific, having fewer workers than \ac{AMT} but providing more fine-grained control; for example, contrary to AMT, Prolific allows screening participants based on their language fluency.

The second independent variable is the gesture generation model used, which is described further in Subsection~\ref{ssec:stimuli}. %. In this experiment we used gesture generation models from Kucherenko et al.~\cite{kucherenko2020gesticulator}. Namely, the models ``No PCA'' (which uses both acoustic features and text as input to generate gestures as an output) and ``No text'' (which uses acoustic features only as an input) were used in this experiment. Those models were chosen as the ``No PCA'' model was rated the highest and the ``No text'' model was rated the lowest in terms of perceived human-likeness ~\cite{kucherenko2020gesticulator}. 
%which we expect should yield a strong preference toward "No PCA" model. 

The two main dependent variables are:

\begin{enumerate}
    \item \textbf{Preference score} \\ 
    whether participants indicate that one video is more human-like than the other, or that they are both equally human-like. Thus, this is a variable with 3 levels.
    From the preference score we also derive inter-rater \ac{ICC} and intra-rater \ac{ICC} scores.
    \item \textbf{Number of attention checks passed} \\
    There are two types of attention checks: \ac{AVAC}, where the participant was instructed to mark a video as broken when they hear or see an instruction to do so; and \ac{SVAC}, where the exact same two videos were played, and the participant was expected to say that there was no difference between the two stimuli.
\end{enumerate}

For exploratory analysis we also consider the following:
\begin{enumerate}
    \item \textbf{Time spent on each rating}
    \\
    The time elapsed from the moment the two videos are shown and the moment the preference is indicated, in seconds.
    \item \textbf{Comment field length}
    \\
    At the end of the experiment participants could leave a free-text comment about the experiment. The length of these comments (in characters) was a measure of their engagement with the experiment. 
\end{enumerate}

We hypothesize that:
\begin{itemize}

  \item \textbf{H1}) Preference for the two models will be significantly different between the results obtained in-lab and the results obtained from Prolific and AMT.
 \item \textbf{H2}) In-lab participants will pass more attention checks during the experiment than online participants.
 %The number of attention tests passed during the experiment will be significantly higher in the results obtained in-lab than the results obtained from online participants.
 \item \textbf{H3}) The inter-rater agreement, estimated using \ac{ICC}, will be significantly higher in the results obtained in-lab than the results obtained from online workers.
 %prolific academic and AMT.
 %\item H4) If there is a significant difference in the preference scores and attention checks in the three testing conditions, then filtering participants based on attention checks removes this difference, and produces a result which is not statistically different.
\end{itemize}
This work was pre-registered using the OSF platform: \href{https://osf.io/dxwak}{osf.io/dxwak}. There are however a few changes made with respect to the pre-registration. The main difference is that the measure for inter-rater agreement was changed to use \ac{ICC} instead of Cohen's kappa. We also removed a fourth hypothesis (H4), since it became irrelevant.

\subsection{Procedure}
A web-based evaluation platform was implemented which was used across all three participant pools. There were some minor differences across each pool to accommodate some differences in how the recruiting was performed. The participants went through the following steps: 
\begin{enumerate}
    \item In-lab: short description and field to input name, \\Prolific: Demographic input for age, gender, employment, and education
    \item Instructions
    \item Five training trials
    \item 21 trials as in Figure \ref{fig:setting}
    \item Demographic questions
\end{enumerate}

 Every participant first completed a training phase to familiarize themselves with the task and interface. This training consisted of five items with video segments not present in the study, showing the participants what kind of videos they may encounter during the study. 
 Then each participant was asked to evaluate 21 same-speech video pairs: 15 pairs randomly sampled from a pool of 28  %random 
 segments and 6 pairs that were intended as attention checks, described in Section \ref{subs:atten_checks}. The videos were presented side by side and could be replayed separately as many times as desired. For each pair, participants indicated which video they thought best corresponded to the given question -- \textit{In which video are the character's movements most human-like?} -- there was also an option to state that they perceived both videos to be equally human-like. An example of how a trial looked like is shown in Figure~\ref{fig:setting}.

The video pairs for each participant were randomly sampled from a pool of videos, while the placement for the \acp{AVAC} was counter-balanced.
Also, the relative position of the videos within each pair (left or right) for each trial was randomized. 
Each participant was randomly assigned to a specific order of videos in the experiment. The same set of 24 fixed orders for trials were used among all the three experimental conditions. For all three conditions we recruited 24 participants to allow for counterbalancing of the order of placement for the \acp{AVAC}. 

\begin{figure}[b]
  \centering
    \includegraphics[width=0.45\textwidth]{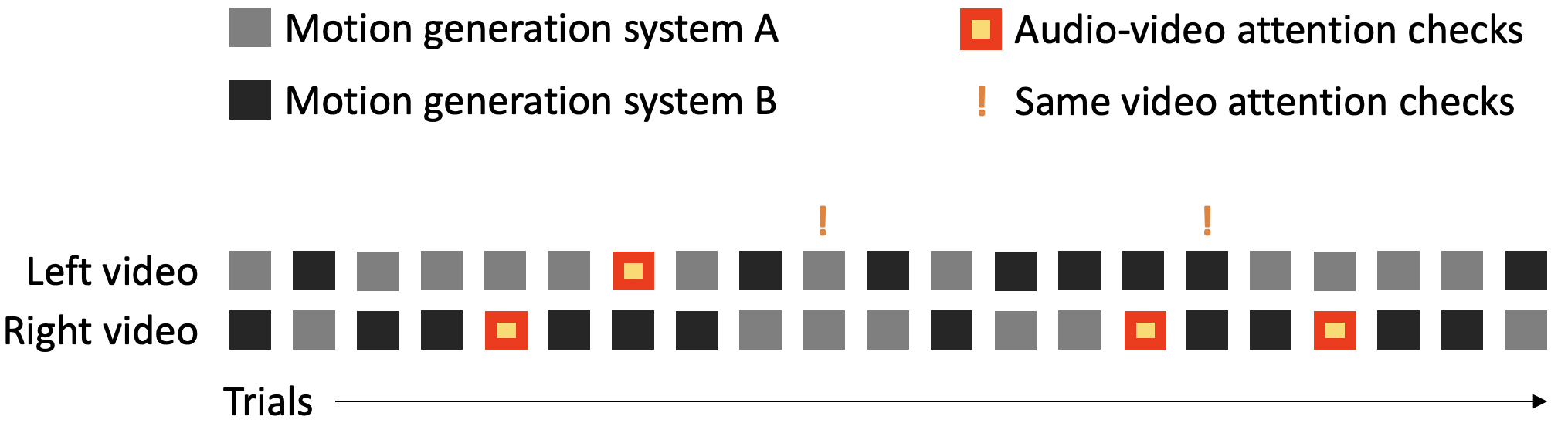}
  \caption{Illustration of attention check placement showing an experimental session for a participant. An experiment consisted of 21 trials, each showing a video on the left and on the right. The videos were showing an avatar generated using gesture generation system A or B or an \ac{AVAC}.}
  \label{fig:trial}
\end{figure}

\subsection{Stimuli}
\label{ssec:stimuli}
The stimuli were generated by gesture generation models from Kucherenko et al.~\cite{kucherenko2020gesticulator}, which are neural networks trained to generate gestures based on speech using a dataset of human gesticulating (see the paper ~\cite{kucherenko2020gesticulator} for more details). For our experiments we have used the following two variations of the model: ``No PCA'' (which uses both acoustic and semantic features as an input and could hence generate complex gestures) and ``No text'' (which uses acoustic features only as an input and hence were mainly generating beat gestures).
Please see samples at: \href{https://vimeo.com/showcase/7571619}{vimeo.com/showcase/7571619}.

\begin{table*}
 \begin{tabular}{llll} 
 \hline
   & \ac{AMT} & In-lab & Prolific \\ 
 \hline
 Number of participants & $24$ & $24$ & $24$ \\ 
 
 Age (avg$\pm{}$std) & $41\pm13$ & $28\pm6$ & $28\pm6$ \\ 
 
 Gender (f/m/o) & 12/11/1 & 10/13/1 & 10/13/1 \\
 
 Exp. with technology (1-5) (avg$\pm{}$std) & $3.8\pm1.0$ & $4.3\pm0.9$ & $4.0\pm0.8$ \\
 
 Number of past similar studies (avg$\pm{}$std) & $0.6\pm2.0$ & $0.8\pm1.4$ & $0.3\pm0.6$ \\
 
 Number of different nationalities & 3 & 16 & 12 \\
 
 Having a higher education\textsuperscript{1} & 58\% & 96\% & 96\% \\
 
 Currently students & 0\% & 66\% & 46\% \\
 \hline

\end{tabular}
\captionsetup{justification=centering}
\caption{Demographics of the participants broken down over the three participant pools. \\
\textsuperscript{1}Higher education is defined as having completed a bachelor's degree or higher}
\label{tab:demographics}
\vspace{-2mm}
\end{table*}
\subsection{Attention checks}
\label{subs:atten_checks}
The attention checks were developed so that they would be similar to the actual task in order to prevent affecting the results \cite{hauser2018manipulation}. For the four \acp{AVAC} we picked four separate video pairs used only for attention check and added either a text or a synthesized voice telling the participant to report the video as broken (two of them had a text, and two had an auditory instruction). These were positioned in one out of four non-overlapping segments (spanning all of the 21 trials) by randomly choosing a place within that segment. The order of attention checks (such as "audio1", "video2", "video1", "audio2") was counterbalanced in a Latin Square fashion. 

The two \acp{SVAC}, which presented the exact same two videos (which were not used for the rest of the study) were placed at the 10th and the 16th trial-position for all experimental sessions. Here, an attentive rater should answer ``no difference''. Figure \ref{fig:trial} illustrates an example of how attention checks were placed within an experiment.

\subsection{Participants}
Participants were recruited from three participant pools: in-lab, Prolific and \ac{AMT}. In total 72 participants were recruited with 24 participants in each participant pool. Since Prolific allows for controlling for a wide range of participant characteristics, it was to a large extent possible to replicate certain demographics of the in-lab study participants (age, gender, education level, student status), see Table~\ref{tab:demographics}. Unfortunately, the same was not possible for \ac{AMT}.
%However, as can be seen in the table, the ``currently students'' demographic is not exactly the same between In-lab and Prolific, due to a discrepancy between what users were requested from Prolific and what they self-reported in our study. 

\subsubsection{In-lab}
In-lab participants were recruited through Facebook posts and the study was performed in person, always using the same laptop in the same room where one of the researchers was present during the whole experiment. Participants could adjust the volume during the training stage.
They received minimal verbal instructions (asked to sit down, put on headphones and follow the on-screen instructions). All the participants were residing in Stockholm, Sweden. No exclusion criteria was used, but the participants were told that they would not get the reward if they failed too many attention checks. This was just used to motivate them to pay attention in the study, in reality everyone would have received their reward; in any case, none of the participants failed too many attention checks. The reward for participation was a movie ticket voucher (average price of a movie ticket in Stockholm is 9.4 USD).

\subsubsection{Prolific}
As the second participant pool, we recruited participants using a crowd-sourcing platform called Prolific based on the same basic demographics as the in-lab study. We requested participants with the same four demographic characteristics as in-lab participants: age, gender, education level and employment status. Payment was 3.90 + 5.5 USD (5.5 USD was given as a bonus upon completion if the participant passed the majority of the audio-video attention checks). When the Prolific participants started our experiment they were asked to fill out a form with the four demographic questions mentioned above. The participants did not know which criteria had given them access to the experiment; and in case they did not match with the criteria provided in their profile, they were not allowed to partake in the experiment.

\subsubsection{Amazon Mechanical Turk}
As the third pool, participants were recruited through the crowd-sourcing platform \ac{AMT}. Payment was 3.90 + 5.5 USD (5.5 USD was given as a bonus upon completion if the participant passed the majority of the audio-video attention checks). Requirements to partake in our experiment was to have finished at least 10,000 previous HITs, have an approval rate of at least 98\% and to be located in the United States.

\ac{AMT} provides a limited way of controlling demographics, thus it was not possible to adequately replicate the demographics of the participants from the other pools.

% \begin{center}

% \end{center}
\section{Results}
The results from the experiment are presented below. The data was analysed using R version 3.6.3. The analysis was performed in a double-blind fashion, such that the authors had obfuscated the participant pool and preference score variables, and revealed them after all analyses were done. Before the comparison of preference score and comparison of attention checks were done, a pre-analysis was conducted in order to determine whether these measures were correlated or not. The outcome of the pre-analysis would determine whether they would be analyzed together (in case they were correlated) or separately (in case they were not correlated).

Participants could mark any video as having issues during the experiment. All trials with videos that were marked as having issues were excluded from the analyses.
In case of missing data, we removed the trial (row) from the participant, but still used the rest of the data from the participant. Therefore, the final number of trials analysed was 359 for Prolific, 360 for AMT, and 358 for in-lab.

% The raw preference score was -1 for the video on the left, 0 if the videos were rated equal, 1 for the video on the right. This score was transformed to be coded as -1 (preference for video based on gesture generation model A), 0 (videos are equal), 1 (preference for video based on gesture generation model B). Any references to ``preference score'' refers to this coding of the data. 

The preference score can have one of 3~possible values: -1 (preference for the `No text' Model), 0 (equal preference) and 1 (preference for the `No PCA' Model). Any references to ``preference score'' refers to this coding of the data. 

\subsection{Pre-analysis}
In order to determine whether to analyze the two dependent variables (DVs) together or separately, we calculated the correlations between the two DVs (average number of \acp{AVAC} passed and preference strength) using the Pearson correlation coefficient. The preference strength was defined as: for each participant, rows with ties (value 0) were removed, so that only values of -1 and 1 remained. Afterwards we calculated the absolute value of the average preference score: $abs(average (preference\_score))$, which we call \textit{preference strength}, which is a continuous variable between 0 and 1. This score indicates how strong the opinion of a user is: the closer the preference is to one, the stronger the preference is to any of the two gesture generation conditions. 
The cut-off for the correlation coefficient $r=0.3$ was determined following standards in Psychology \cite{cohen2013statistical}. It was found that for all the conditions the correlation coefficient $r$ was lower than the cut-off value. Hence we regarded the two variables as not correlated and analyzed them separately. 
The preference strength was only used for the pre-analysis, for all the comparisons described below the preference score was used.

% For the calculation of correlation between preference score and number of attention checks passed, 

%, while a value of zero would indicate either choosing both conditions equally often, or choosing that there is no difference between them. 

%1) rows with ties (value 0) were removed, so that only values of -1 and 1 remained;
%2) we calculated the average preference score, by summing all the -1 and 1 scores and dividing them by the number of trials. This was also a value between -1 and 1;
%3) we took the absolute value of this average. In this way, we obtained a “preference strength”: the closer this value is to 1, the more the participant showed a preference towards a certain video; the closer this value is to 0, the more the participant showed no consistent preference.

% \begin{table}[]
%  \begin{tabular}{llll} 
%  \hline
%   & In-lab & Prolific & AMT \\ 
%  \hline
%  passed audio attention checks & 100\% & 100\% & 100\% \\ 
%  \hline
%  passed video attention checks & 97\% & 100\% & 100\% \\ 
%  \hline
%  passed same video attention checks & 69\% & 79\% & 71\% \\ 
%  \hline
% \end{tabular}
% \caption{}
% \end{table}

\begin{figure}
  \centering
    \includegraphics[width=0.4\textwidth]{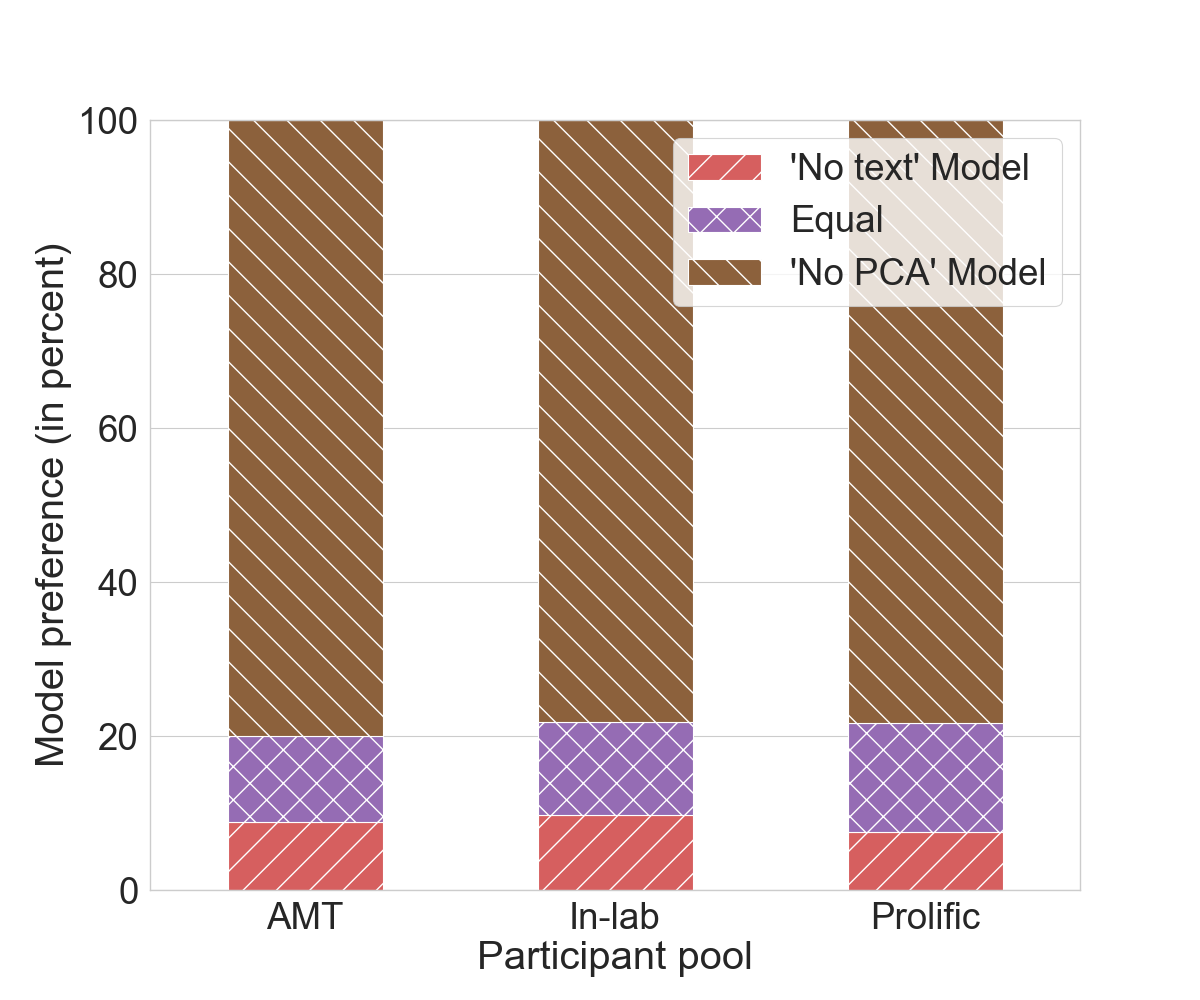}
    \vspace{-3mm}
  \caption{Participants' preferences toward the two different gesture generation models per participant pool.}
  \label{fig:perception_scores}
\end{figure}

\subsection{Comparison of preference score}
We fitted a cumulative link mixed model via likelihood ratio test with preference score as dependent variable, participant pool as predictor, and rater and sample (refers to a unique pair of videos) as random intercepts; however, we found no significant effect of participant pool ($ \chi ^2(2)=0.54, p=0.77$). A post-hoc g*power analysis was made, resulting in a power of 0.95.

\begin{figure}
  \centering
    \includegraphics[width=0.4\textwidth]{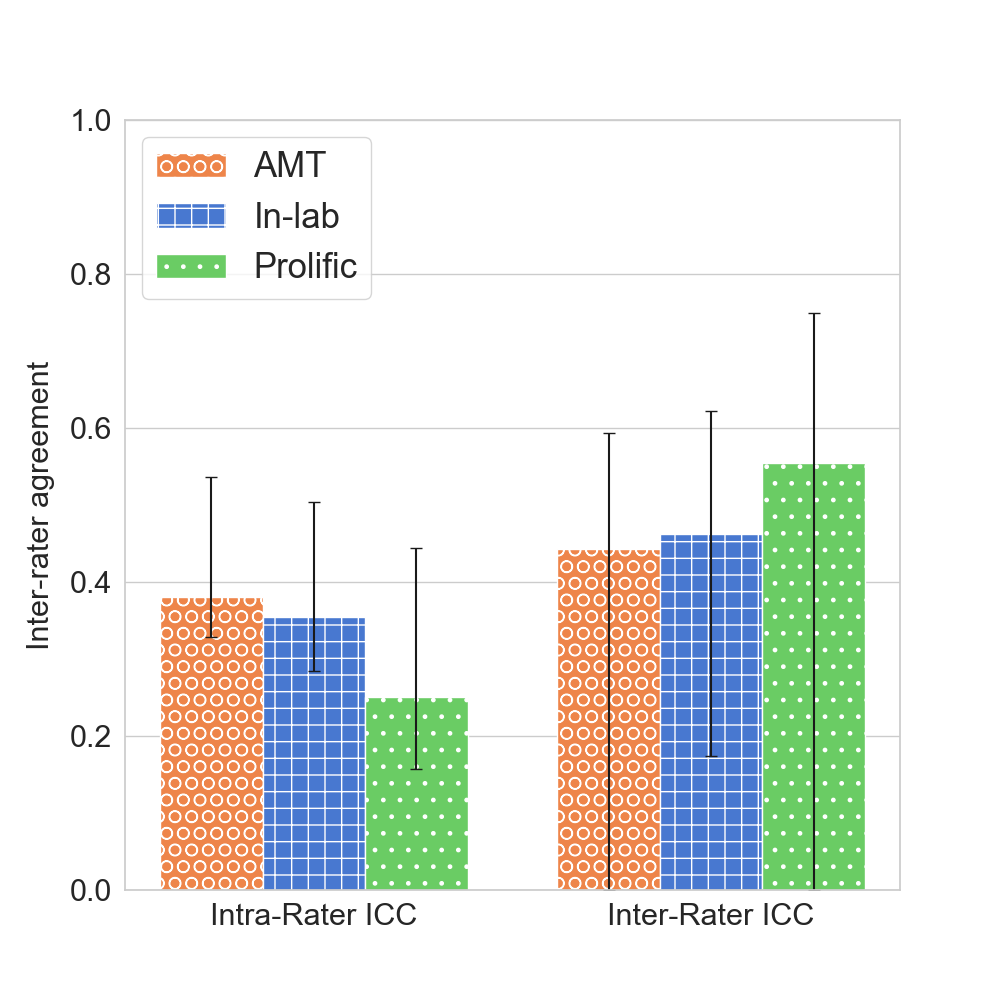}
  \caption{Inter-rater and Intra-rater \ac{ICC} for each model separately broken down per participant pool. Higher values corresponds with higher agreement between the raters.  Error bars show 95\% confidence intervals.}
  \label{fig:inter-rater-v2}
\end{figure}

As a separate test, we compared inter-rater and intra-rater reliability. An analysis using bootstrapped ICC  was used for the inter-rater and intra-rater agreement using the ``agreement'' R-package \cite{agreement} and the dimensional ICC method using Model 2A \cite{gwet2014handbook}. Analysis showed that the confidence intervals are overlapping (as seen in Figure~\ref{fig:inter-rater-v2}) and that there was no statistically significant difference between the three participant pools in either of the cases \cite{stolarova2014assess}. %Table~\ref{tab:inter-intra-rater}.% and .

%  An analysis using bootstrapped ICC was used for the inter-rater agreement calculation. Since the distribution of responses was heavily skewed toward the ``No PCA'' model, we analyzed each preference score separately. The analysis showed that there was no statistical significant difference between the participant pools for none of the models, see Fig~\ref{fig:inter-rater}. Raters agreed more with each other when rating the ``No PCA'' model to be considered more human-like as opposed to when they chose the ``No text'' model or deeming both models equal.
 
%  \subsubsection{Intra-rater agreement}
%  An analysis using bootstrapped ICC was used for the intra-rater agreement calculation and showed no statistical significant difference between the three participant pools, see Table~\ref{tab:inter-intra-rater} and Figure~\ref{fig:inter-rater-v2}

%We will calculate Cohen's Kappa for the preference score for all three conditions and compare if they are different by checking if confidence intervals (α = 0.05) are overlapping (Stolarova et al. 2014).
% X_2004011349 = 0.54 (0, 0.708)
% Y_2004011349 = 0.571 (0, 0.734)
% Z_2004011349 = 0.687 (0, 0.84)

\subsection{Comparison of passed attention checks}
\label{sec:compare_att}
\subsubsection{Audio-video attention checks} The \acp{AVAC} were coded as 0 (failed) or 1 (passed). Then for each participant, we calculated the average attention score, by summing all the 0 and 1 scores and dividing them by the number of attention checks. This was a value between 0 (high failure rate) and 1 (low failure rate). Most of the participants (69 out of 72) passed all audio-video attention checks during the experiment and thus there was no difference between the participant pools in terms of passing \acp{AVAC}. The three participants that did not pass all of the \acp{AVAC} failed only on the video-based attention checks and belonged to the in-lab participant pool. Therefore we concentrated our analyses on the \acp{SVAC}.

\subsubsection{Same video attention checks}
\begin{figure}
  \centering
    \includegraphics[width=0.32\textwidth]{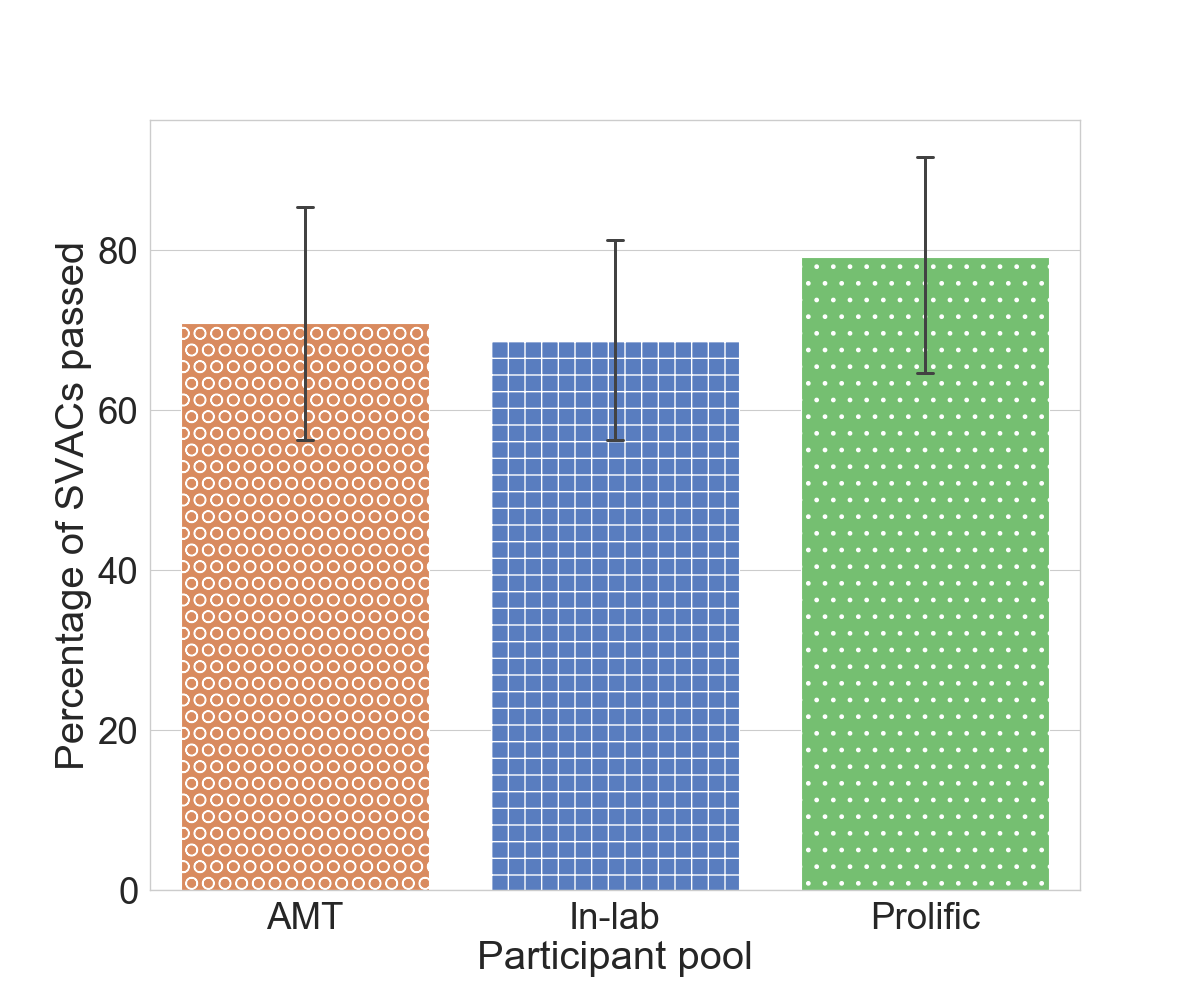}
  \caption{Percentage of passed \acs{SVAC} broken down per participant pool. A higher number indicates higher attention. Error bars show 95\% confidence intervals.}
  \label{fig:att_checks}
\end{figure}

The \acp{SVAC} were coded as 0 (failed) or 1 (passed). Then for each participant, we calculated the average of passed attention check as in Section~\ref{sec:compare_att}. Compared to the \acp{AVAC}, a higher degree of participants (32 out of 72) failed either one or both of the \acp{SVAC}. The results contained a large number of zero-values, we therefore fit a zero-inflated regression model on the same \acp{SVAC} via maximum likelihood estimation ($ \chi ^2(2)=0.17, p=0.91$) and observe that there is no statistically significant difference between the participant pools. These results are visualized in Figure \ref{fig:att_checks}.

% \begin{figure}
%   \centering
%     \includegraphics[width=0.5\textwidth]{figures/plot_inter_rater_agreement.png}
%   \caption{Inter-rater agreement for each model separately broken down per participant pool. Higher values corresponds with higher agreement between the raters.}
%   \label{fig:inter-rater}
% \end{figure}

\iffalse
\begin{table}
\begin{tabular}{lllll}
\hline
 Participant pool & Intra-Rater ICC & Inter-Rater ICC  \\
\hline
AMT              & 0.380 (0.329, 0.537) & 0.443 (0.000, 0.594) \\
In-lab           & 0.354 (0.285, 0.504) & 0.462 (0.174, 0.622) \\
Prolific         & 0.251 (0.157, 0.444) & 0.554 (0.000, 0.749) \\  
\hline
\end{tabular}
\caption{Alternative table for inter/intra rater agreement}
\label{tab:inter-intra-rater}
\end{table}
\fi

% \begin{table*}[]
% \begin{tabular}{llllll}
% \hline
%  Participant pool  & `No text' model  & Equal & `No PCA' model  \\
% \hline
% AMT              & 0.152 (0.053, 0.227) & 0.123 (0.072, 0.162) & 0.800 (0.750, 0.843) \\
% In-lab           & 0.105 (0.025, 0.171) & 0.140 (0.087, 0.196) & 0.796 (0.746, 0.840) \\
% Prolific         & 0.111 (0.072, 0.143) & 0.187 (0.089, 0.248) & 0.796 (0.730, 0.850) \\  
% \hline
% \end{tabular}
% \end{table*}

\subsection{Trial duration analysis}

\begin{figure}
  \centering
    \includegraphics[width=0.33\textwidth]{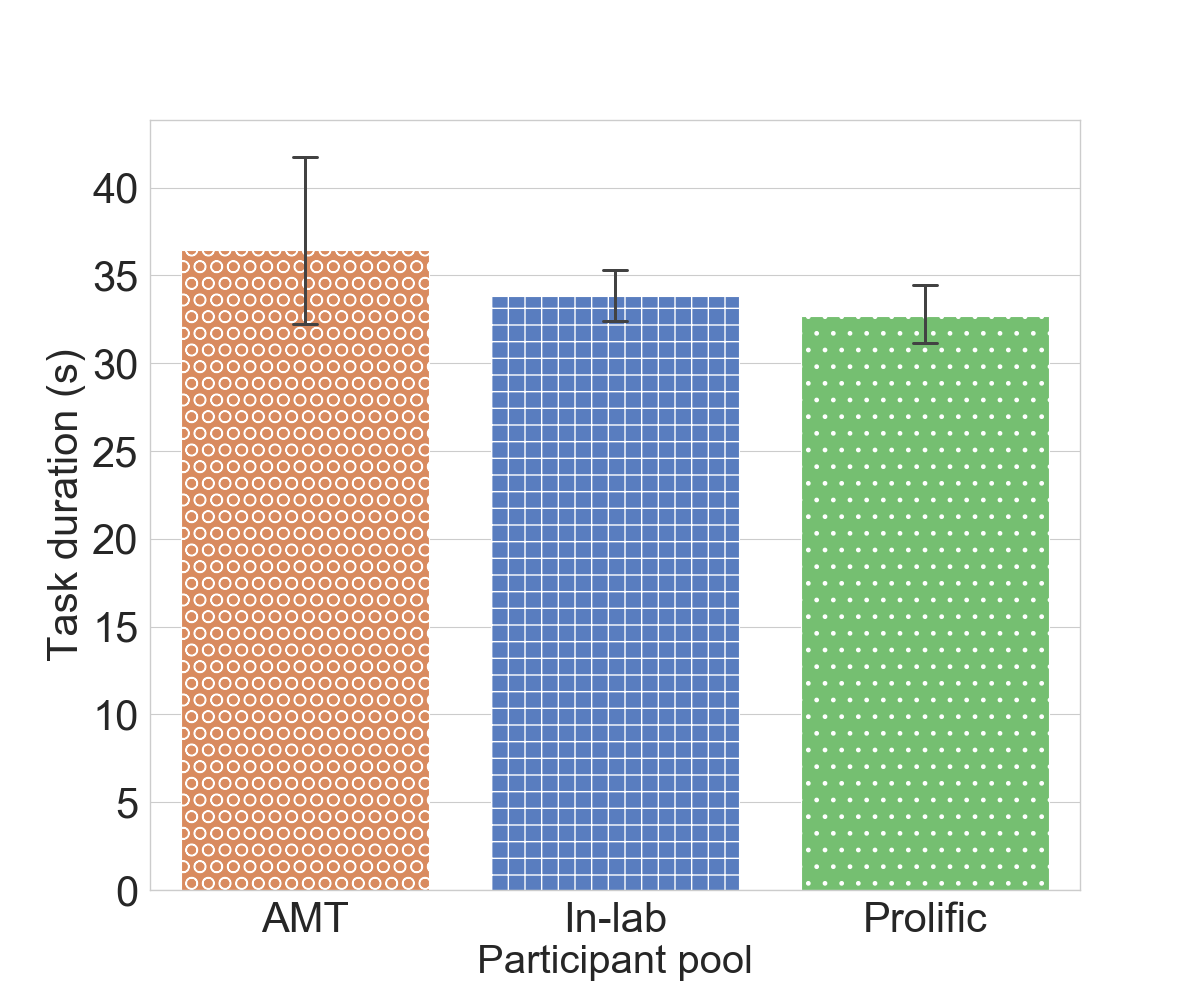}
  \caption{Average trial duration in seconds broken down per participant pool. Error bars show 95\% confidence intervals.}
  \label{fig:task_dur}
\end{figure}

As an exploratory analysis we considered the difference between the participant pools in terms of the duration to complete each trial, since it can be used to measure attentiveness~\cite{curran2016methods}. We performed the analysis by fitting a linear mixed-effects model using the lmerTest package \cite{kuznetsova2017lmertest}, with task duration as dependent variable, participant pool as predictor, and participant and sample as random intercepts.
We found no statistical significance between the participant pools ($ \chi ^2 (2) = 1.23, p = 0.54$), meaning that participants spent approximately the same amount of time on the trials in each participant pool. The \ac{AMT} participant pool showed a higher variance than the two other participant pools as can be seen by the larger confidence interval in Figure~\ref{fig:task_dur}.

\subsection{Comment field analysis}

\begin{figure}
  \centering
    \includegraphics[width=0.3\textwidth]{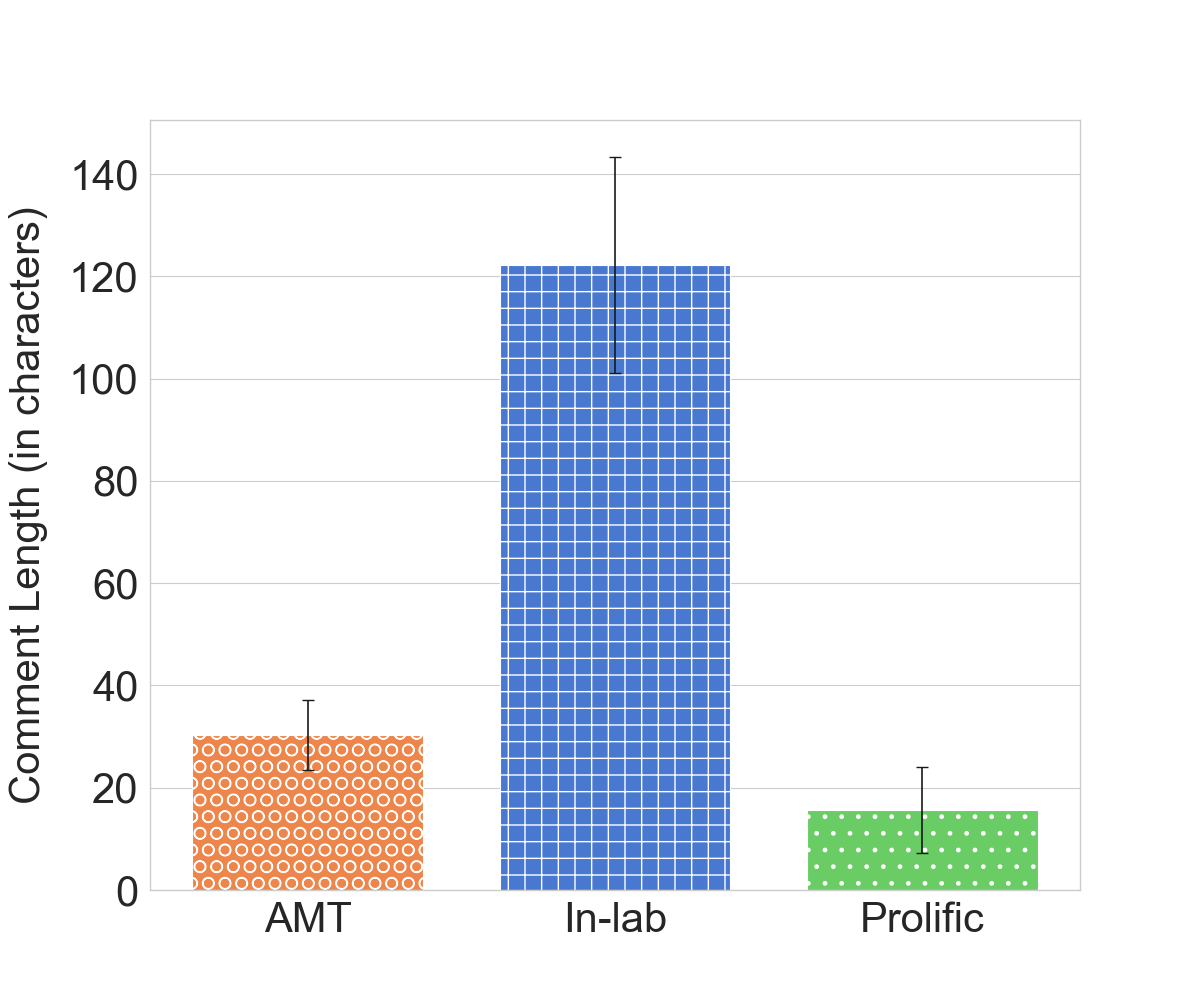}
  \caption{Average comment length broken down per participant pool. Error bars show 95\% confidence intervals.}
  \label{fig:res_comments}
  \vspace{-2mm}
\end{figure}

As an additional exploratory analysis, we also looked at the length (in characters) of the optional comment at the end of the experiments as it might reveal information on how engaged participants in the different pools were.

Since several participants left the field empty, thus resulting in a value of 0, we fit a zero-inflated regression model on the comment length via maximum likelihood estimation, with participant pool as predictor. This model was significant, suggesting that participants' comments differed in the three participant pools ($ \chi ^2 (2) = 800.99, p < .001$). As a post-hoc analysis we calculated the estimated marginal means (EMMs) using Tukey correction for multiple comparisons. The results are illustrated in Figure \ref{fig:res_comments}.  We found that in-lab participants wrote significantly longer comments than both AMT participants ($z = 4.120, p < .001$) and Prolific ($z = 4.671, p < .001$), while there was no statistically significant difference between AMT and Prolific ($z=1.364, p = 0.36$).

\section{Discussion}
We compared three participant pools; in-lab, Prolific, and \ac{AMT} in terms of their preference scores, inter/intra-rater agreement, and attentiveness when comparing two gesture generation models. 

From the results we see that there is no difference in neither of the three measures across the three participant pools, thus we reject the three hypothesis we set out from the beginning; H1, H2, and H3. Our result is  consistent with previous findings on other perceptual tasks \cite{germine2012web, woods2015conducting, lansford2016use}. The preference scores obtained across the three participant pools were also consistent with the experiment performed in the work of Kucherenko et al. \cite{kucherenko2020gesticulator}. We therefore conclude that reliable results can be obtained from both in-lab participants and online workers. 

Comparing the preference score alone might not give a complete picture of the differences between the two groups and thus we also investigated the inter- and intra-rater agreement. The Inter-rater agreement gives a measure of how consistent an individual participant is with the other participants, while the intra-rater agreement gives a measure of how consistent each worker are with themselves. We compared these measures as well, and found that participants are equally consistent in both of these regards over all of the three participant pools. This supports our conclusion that the participants in all conditions were equally reliable.

In terms of the number of passed attention checks there was however a large difference between our experiment and that of Kucherenko et al~\cite{kucherenko2020gesticulator}. In the previous study a big portion of the \ac{AMT} workers (over 75\%) did not finish the study, either due to timing out or failing a majority (more or equal to two) of the \acp{AVAC}, while in the present study participants from the \ac{AMT} pool (as well as those from the other pools) never failed more than one \ac{AVAC}, and no one timed out. There were two major differences between Kucherenko et al.'s study and the present one; how the \acp{AVAC} were designed and the size of the reimbursement. The \acp{AVAC} in the current work explicitly either displayed a text message or spoke using a synthetic voice instructing the user to mark the trial as broken. In Kucherenko et al.'s work the \acp{AVAC} were not explicit, instead the quality of the audio or video was degraded to such a level where they were unusable, and participants were asked to report any broken videos that prevented them from making a judgement. These implicit \acp{AVAC} probably led to more participants failing them. They might even have been a cause of frustration for participants, thus decreasing their intrinsic motivation to take part in the task. The second main difference between this experiment and that of Kucherenko et al. lies in the monetary reward, which was considerably higher in the present study (28.2 USD/h vs 9 USD/h). This seems to suggest that the reimbursement level can have a strong effect on how attentive the participants were and that when having high reward \acp{AVAC} might not be necessary, as participants might be more motivated to perform well. Giving a more appealing monetary reward might have increased participants' extrinsic motivation to complete the task. Both intrinsic and extrinsic motivation are important for collecting high quality data \cite{kaufmann2011more}. 

Another point of interest is that the in-lab participants provided longer comments in the optional comment field of the experiment, which might suggest that in-lab participants put more effort and commitment into the task. Participants did not differ significantly in terms of time spent on each trial, however the \ac{AMT} did show a considerably higher variance. This is interesting and warrants further investigation.

%Thus, our results suggest that %by carefully tailoring the attention checks to the task, and 
%by providing appropriate monetary reward, perceptual studies can be conducted using %online participants and obtain the same results as in-lab.

% We tried to control the demographics on \ac{AMT} via filtering participants by ourselves. We asked \ac{AMT} workers to provide their demographics in order to obtain \textit{qualification} to do the task. The idea was to give the \textit{qualification} only to those workers which fit into the desired demographics. This did not work because of two reasons: 1) \ac{AMT} workers were sharing with each other "correct" demographics cheating the filter; 2) There were very few students among \ac{AMT} workers.
% Because of these issues we decided to not restrict demographics of \ac{AMT} workers.

\vspace{-1mm}
\subsection{Limitations}
The results from our study seem promising for researchers who want to use crowd-sourcing as a method for evaluating stimuli and performing perceptual experiments. However, there are two main points which the authors would like to highlight as limitations to the current experiment.

In this study we aimed at reimbursing the participants in an as comparable way as possible. Due to restrictions by the University it was not possible to reimburse in-lab participants monetarily. Therefore each participant was given a cinema ticket voucher upon successful completion of the study. The participants on Prolific and \ac{AMT} were given the monetary value of the cinema ticket (approx. 9 USD). The reimbursement of 9 USD for 20min task is relatively high for Prolific and AMT and, as previously discussed, could be an important factor when comparing with previous work \cite{kaufmann2011more}.

The reimbursement on both \ac{AMT} and Prolific was divided in two steps, one for finishing the study, and a second part for finishing the study ``successfully''. The wording was intentionally ambiguous to not give away that there were attention checks, and all participants who failed several \acp{AVAC} would not have been reimbursed. The two levels were set up as we wanted to reimburse even ``cheating'' participants since we intended to use their data, but we also did not want to pay ``cheaters'' the full amount. The in-lab participants were informed that they would not receive the reward if they did not finish the experiment successfully (however, they would still receive it even if they failed). These differences could potentially be confounds.

\subsection{Future Work}
Many previous experiments have used lower reimbursement levels compared to the present study, and has reported high failure rates for attention checks ~\cite{jonell2019learning, yoon2019robots, kucherenko2020gesticulator}. In the present study none of the participants failed more than one \ac{AVAC}, which seems to indicate that reimbursement might be an important factor. As a next step we plan to analyze the effect on the results in a similar audio-visual perception test by varying the reimbursement levels.

Furthermore it would be interesting to investigate the influence of different types of attention checks that could be used in this type of perceptual experiments and how they would affect the results.

Another interesting direction for future work would be to conduct the same experiment with \textit{experts}, i.e. people who work professionally in the field of gesture analysis, in order to validate the finding based on ``correct'' answers from experts.
\section{Conclusion}
This paper presented an experiment where a comparison was made between three participant pools; in-lab, Prolific, and Amazon Mechanical Turk. The experiment was a subjective preference test of videos generated by two gesture generation models. The results showed that there was no significant difference in several measures (preference score, attention checks passed and inter/intra-rater agreement) between the three participant pools. These results indicate that online workers can successfully be used instead of in-lab participants for audio-visual perception experiments similar to the one outlined in this paper, significantly easing the task of recruiting participants. The results have to be interpreted with some caution however, as the effect of reimbursement level is not fully understood, and would need further investigation.

% 
% compared to previous work the reimbursement level could have a significant impact on the quality of the results and should be investigated further.%, and the fact that the reimbursements weren't identical across the participant polls in the presented study. 

% These results indicate that online workers are as attentive as in-lab participants for experiments for  provided they receive the same reimbursement. %This might be due to the high reimbursement, so future work should look into whether the amount of reimbursement has a significant impact on response quality.

\section*{Acknowledgement}
% The authors would like to thank ** for helpful discussions. 
This work was partially supported by the Swedish Foundation for Strategic Research Grant No.: RIT15-0107 (EACare), and by a WASP Expedition Project on Correct by-design and Socially Acceptable Autonomy (CorSA).

%\begin{acks}
%\end{acks}

\bibliographystyle{ACM-Reference-Format}
\bibliography{sample-base}

\end{document}